\documentclass[conference]{IEEEtran}

\usepackage{url}
\usepackage{cite}
\usepackage{xspace}
\usepackage{multirow}
\usepackage{ifthen}
\usepackage{xcolor,colortbl}
\usepackage{tabulary}
\usepackage{balance}
\usepackage{booktabs}
\usepackage{graphicx}
\usepackage[caption=false,font=footnotesize]{subfig}

\newboolean{authnotes}
\setboolean{authnotes}{false}

\newcommand{\camera}[1]{{#1}}

\ifthenelse{\boolean{authnotes}}
{
\newcommand{\usman}[1]{\footnote{{\bf Usman: #1}}}
\newcommand{\parag}[1]{\footnote{{\bf Parag: #1}}}
\newcommand{\alex}[1]{\footnote{{\bf Alex: #1}}}
\newcommand{\memo}[1]{{\color{red} #1}}
\usepackage{draftwatermark}
\SetWatermarkText{}
}
{
\newcommand{\usman}[1]{}
\newcommand{\parag}[1]{}
\newcommand{\alex}[1]{}
\usepackage{draftwatermark}
\SetWatermarkText{}
\newcommand{\memo}[1]{#1}
}

\newcommand{\finding}[2]{\vspace{2mm}\noindent \textbf{Finding \##1: #2\vspace{2mm}}}

\newcommand{\lora}{LoRa\xspace}
\newcommand{\lorawan}{{LoRaWAN}\xspace}

\makeatletter
\define@key{Gin}{Trim}
           {\let\Gin@viewport@code\Gin@trim\expandafter\Gread@parse@vp#1 \\}
\makeatother

\begin{document}
\title{Does Bidirectional Traffic Do More Harm Than Good in LoRaWAN Based LPWA Networks?}

\author{\IEEEauthorblockN{Alexandru-Ioan Pop\IEEEauthorrefmark{1}\IEEEauthorrefmark{2}, Usman Raza\IEEEauthorrefmark{2}, Parag Kulkarni\IEEEauthorrefmark{2}, Mahesh Sooriyabandara\IEEEauthorrefmark{2}}
\IEEEauthorblockA{\IEEEauthorrefmark{1}  Faculty of Engineering, University of Bristol, Bristol, UK\\
Email: ap14621@my.bristol.ac.uk}
\IEEEauthorblockA{\IEEEauthorrefmark{2}Telecommunications Research Laboratory, 
Toshiba Research Europe Limited, Bristol, UK\\
Email: firstname.lastname@toshiba-trel.com}
}

\setlength{\columnsep}{0.23in}


%

\maketitle
\begin{abstract}


The need for low power, long range and low cost connectivity to meet the requirements of IoT applications has led to the emergence of Low Power Wide Area (LPWA) networking technologies. The promise of these technologies to wirelessly connect massive numbers of geographically dispersed devices at a low cost continues to attract a great deal of attention in the academic and commercial communities. Several rollouts are already underway even though the performance of these technologies is yet to be fully understood. In light of these developments, tools to carry out `what-if analyses' and pre-deployment studies are needed to understand the implications of choices that are made at design time. While there are several promising technologies in the LPWA space, this paper specifically focuses on the LoRa/LoRaWAN technology. In particular, we present LoRaWANSim, a simulator which extends the LoRaSim tool to add support for the LoRaWAN MAC protocol, which employs bidirectional communication. This is a salient feature not available in any other \lora simulator. Subsequently, we provide vital insights into the performance of LoRaWAN based networks through extensive simulations. In particular, we show that the achievable network capacity reported in earlier studies is quite optimistic. The introduction of downlink traffic can have a significant impact on the uplink throughput. The number of transmit attempts recommended in the LoRaWAN specification may not always be the best choice. We also highlight the energy consumption versus reliability trade-offs associated with the choice of number of retransmission attempts.

\end{abstract}
\IEEEpeerreviewmaketitle

\section{Introduction}

The number of connected devices has already surpassed the number of inhabitants on this planet and this number is likely to grow significantly moving forward further fueling the Internet of Things (IoT) revolution. 
Many current IoT applications are characterized by low cost sensors with constrained battery life, some of which may be located in remote (perhaps, hostile) locations, reporting small amount of data at a time (low throughput requirement) over a communication link to a backend system. \memo{Deployment constraints such as the sheer volume of these devices, the need to keep costs down, and difficulty/inability to change batteries make it highly desirable that the solution should cater to several years of (literally) zero touch operation}. A key aspect of any such solution is the communication mechanism employed to ferry the data. Given the inadequacies of legacy communication solutions (expensive hardware and communication costs and short battery life), a new breed of connectivity solutions have emerged popularly referred to as \emph{Low Power Wide Area (LPWA)} networking technologies~\cite{razacomst}. \memo{Some of the most prominent ones are \lorawan, SIGFOX, Ingenu RPMA, and Weightless standards}.

\memo{LPWA technologies promise} to wirelessly connect massive number of geographically dispersed devices and sensors at a low cost. With a claimed range of several kilometers and a battery life running into several years while keeping costs down, LPWA technologies have become a promising alternative with large scale deployments already underway in several geographies.

\memo{\lorawan is one of the most popular LPWA technologies, with a growing deployment footprint.} In addition, it is relatively easy to procure and deploy LoRaWAN hardware for experimental purposes which has resulted in LoRaWAN receiving the most attention (in comparison to other technologies) from the research community. The broader objective of this paper is to further the understanding of this technology.

Based on our experiences, we believe that a key limitation that impedes gathering greater insight into LoRa/LoRaWAN technology is the lack of a \memo{comprehensive} simulation tool. Existing simulators~\cite{lancasterlorascale, MaartenWeyn, danishmastersthesis} only allow studying performance in scenarios where there is traffic only in the uplink. In this work, we take this a step forward by presenting LoRaWANSim, which extends an existing simulator~\cite{lancasterlorascale} to support the LoRaWAN MAC protocol, which requires and facilitates bidirectional communication. Having the capability to transmit in the downlink is absolutely crucial because a) it may be \memo{a key requirement for many IoT} applications b) various network management functions of LoRaWAN such as handshaking, network joining, exchange of security keys etc. cannot be accomplished without it, c) adapting communication parameters to get optimal network performance requires feedback from the gateway to the end devices, e.g., the gateway monitors the uplink signal quality so as to inform the end devices that they should adapt their radio parameters, and d) many \memo{Internet} protocols require it (notable recent efforts have been made to extend protocols such as IPv6 to work on constrained devices and networks~\cite{ip6overlora}). All in all, LoRaWANSim could be a handy tool to assess the performance of the technology in a wide variety of scenarios and uncover \memo{useful insights to guide design choices}.

In this paper we make the following contributions:

\begin{itemize}
\item We first present LoRaWANSim, a discrete event simulator that introduces some essential MAC layer features for realistic performance analysis of \lorawan. LoRaWANSim implements downlink traffic, acknowledgments, retransmissions, \memo{and loss-based data rate adaptation schemes, while taking into account the regulatory duty cycle limitations}.
\item By using LoRaWANSim, we then analyze LoRaWAN scalability under more realistic traffic scenarios where dominant uplink traffic is accompanied by some fractional downlink traffic. Findings from this study reveal that \lorawan is negatively affected not only by the network scale as observed in previous studies~\cite{lancasterlorascale}, but also by downlink traffic (data and ACK) and retransmission attempts, which are a major contributor to the reduction in throughput and reliability.
\item We highlight that a careful \memo{choice of} network size and \lorawan parameters can provide reasonable performance. To guide adopters of this technology, we provide a detailed discussion on the effect of different MAC layer parameters (such as number of retransmissions) on energy efficiency, reliability and highlight trade-offs and how LoRaWANSim can be exploited to uncover such useful insights.
\end{itemize}

The rest of the paper is organized as follows: Section~\ref{sec:loraoverview} presents a short primer on \lora and \lorawan. Section~\ref{sec:relwork} discusses related work, while Section~\ref{sec:lorawansim} describes our simulator and the new MAC layer features incorporated in it. In Section~\ref{sec:eval}, we put our simulator into action and analyze the performance of \lorawan for bidirectional traffic, deriving key insights, which are then summarized in Section~\ref{sec:summary}. Finally, Section~\ref{sec:conclusion} concludes this paper.

\section{An Overview of LoRa and LoRaWAN}
\label{sec:loraoverview}
In this section, we \memo{briefly present the \lora LPWA physical layer and the \lorawan MAC layer running on top of it.}
\subsection{LoRa}
\label{subsec:lora}

LoRa is a proprietary wireless physical layer developed by Semtech Corporation. It employs a novel Chirp Spread Spectrum (CSS) modulation technique. LoRa gives the system a large link budget and processing gain, allowing it to decode signal powers below the noise floor, while making it immune to multipath fading, Doppler Shift, and narrowband interference. The range and energy consumption of LoRa depend on multiple physical layer parameters including Spreading Factor (SF), Bandwidth (BW) and Coding Rate (CR). Higher values of SF spread the signal more in time \memo{in order to} put more energy per transmitted bit, allowing successful reception at longer distances. Nevertheless, this also increases the time on air, reducing the effective data rate. Interestingly, different SFs are orthogonal (much like frequency channels) and do not collide even if the LoRa devices use them at the same time.
Large channel bandwidths achieve higher data rates but experience more noise, limiting the range. Furthermore, LoRa can employ Forward Error Correction (FEC) to increase reliability and range. The redundancy added to the transmissions is defined by the CR and slightly increases the time on air. 

\subsection{LoRaWAN}
LoRaWAN is an open protocol standard developed on top of \lora by the LoRa Alliance. It specifies how the end devices should connect to one or more always-on \lora gateways using the unlicensed radio spectrum in the Sub-GHz Industrial, Scientific and Medical (ISM) bands. The gateways are then connected through the backend to network servers and application servers. As regulations on the ISM bands vary across multiple regions, \lorawan makes separate recommendations for Europe, North America, Asia, etc. We restrict our discussion to Europe, given the space constraints. The devices transmit using a simple ALOHA based multiple access scheme and do not employ any carrier sensing or Listen Before Talk (LBT) scheme. Without such scheme, the European Sub-GHz ISM bands (868 MHz and 433 MHz) require the devices to respect a strict duty cycle (1\% is typical for default sub-bands). Furthermore, the transmission power can be set to a maximum of 20 dBm, although most sub-bands limit the power to 14 dBm only. The available spectrum and the regulations have implications on possible values for communication parameters of LoRa (SF, BW and CR). These led \lorawan to recommend a few distinct values for SF (\{7, 8, 9, 10, 11, 12\}), BW (\{125 kHz, 250 kHz, 500 kHz\}) and CR. These values allow the network to make trade-offs between range, data rate and energy consumption as described in \ref{subsec:lora}. 
These values can be decided at design time or changed at run-time using the Adaptive Data Rate (ADR) algorithm of \lorawan, which typically requires the network server and end devices to monitor the quality of the uplink and update parameters to improve reliability and energy consumption. 

LoRaWAN specifies three device classes to cater to different types of applications. Class A devices, the default class, can only receive downlink data immediately following an uplink transmission. Class B devices can wake up periodically to receive scheduled downlink data traffic. Class C devices listen continuously and are typically mains-powered devices. 

To protect the data, \lorawan provides separate network-level and application-level security mechanisms based on symmetric encryption techniques using the private keys.

\section{Related Work}
\label{sec:relwork}

\memo{As we present a simulator for \lorawan LPWA networks, we briefly discuss related tools and studies on \lorawan}.

\memo{\subsection{\lorawan Analytical Models and Simulators}}
Multiple analytical models~\cite{trllorascale, markovAnalyticalMACModel, classbmodel} and simulators~\cite{lancasterlorascale, MaartenWeyn, danishmastersthesis} have been proposed to understand the performance of \lorawan.
None of these models is provide any insights on the interplay between downlink traffic and the gateway's duty cycle limit or effect of MAC parameter settings on the \memo{reliability of \lorawan}. To bridge this gap, we design LoRaWANSim, which extends the functionalities of LoRaSim~\cite{lancasterlorascale}, an existing discrete event simulator. \memo{Other} simulators~\cite{MaartenWeyn, danishmastersthesis} including LoRaSim \memo{focus more on} \lora physical \memo{layer aspects, including modulation, channel effects, and path loss}. Unfortunately, their MAC layer capabilities are very much limited to an implementation of the ALOHA protocol. With LoRaWANSim, we take an important step forward by incorporating \memo{multiple} MAC layer features that are part of the \lorawan standard. These features include the possibility to send downlink traffic, special control messages, confirmed messages, acknowledgments, and retransmissions. By doing so, LoRaWANSim enables users to evaluate the performance of the \lorawan MAC layer, derive useful insights about the effect of several MAC layer parameters, and evaluate possible enhancements to the \lorawan standard.  


\subsection{Performance of \lora and \lorawan}
The performance of \lora based networks has been studied in multiple recent papers~\cite{trllorascale, lancasterlorascale, understandinglimitsoflora, capacityscalability}. With the assumption that LPWA networks are expected to serve a massive number of end devices, most of these studies~\cite{lancasterlorascale, trllorascale, capacityscalability} raise the question if \lorawan would scale well and if so what are the fundamental limitations on network capacity and reliability. It is unveiled that an uncoordinated and random selection of communication parameters (especially SF and channels) by end devices while using an ALOHA protocol causes significant interference~\cite{trllorascale}, limiting network capacity in the presence of large numbers of contending end devices. As a result, a \lorawan gateway is estimated to serve 120 end devices per 3.8 ha \memo{in the worst case}~\cite{lancasterlorascale}\alex{this is the worst case, which we know is unrealistic}, a density far too low for a smart city deployment in an urban environment.
None of the studies mentioned above considers the combined effect \memo{on overall network capacity of downlink traffic and duty cycle limits on the gateway}. To the best of our knowledge, we are the first to provide necessary support in a simulator to study these MAC layer performance aspects.

\section{LoRaWANSim}
\label{sec:lorawansim}

In this section, we present LoRaWANSim, a simulator that extends LoRaSim by introducing support for the LoRaWAN MAC. In particular, this facilitates sending data/ACK messages in the downlink and retransmissions at the end nodes complying with the LoRaWAN MAC protocol. We first briefly describe LoRaSim before discussing additional MAC features available in LoRaWANSim. 

\subsection{LoRaSim}
LoRaSim is a discrete event simulator, which is described in more detail in~\cite{lancasterlorascale}. In LoRaSim, a \lora network can be deployed by placing multiple gateways and end devices randomly in a two dimensional space. With only uplink and no downlink support, end devices generate traffic periodically to be sent to the gateway. For these transmissions, devices can be configured to use any possible value of SF, frequency channel, bandwidth, coding rate, and transmission power. Successful reception of these uplink packets depends on multiple factors in addition to the transmission communication parameters. These include distance dependent path loss, fading, network collisions and receiver sensitivity of the \lora devices. For the first two factors, LoRaSim uses the log distance path loss model based on an empirically measured reference path loss value of 127.47 dB at 40 m distance, a path loss exponent of 2.08 and normal shadowing distribution with zero mean and variance of 3.57\usman{we'll come back to this}. All these values are obtained from characterization of a real environment described in more detail in~\cite{lancasterlorascale}. LoRaSim implements a realistic collision model, where two retransmissions will not interfere if they use different frequencies or spreading factors. It also takes into account the sensitivity of the radio hardware and the capture effect of LoRa \memo{(the ability to receive the strongest signal despite possible collisions with other weaker signals)}.
\subsection{LoRaWANSim: Additional MAC Features} 

\begin{table*}
\caption{Functions of Downlink Messages: the downlink traffic is shown in \emph{italics}}
\label{tab:bidirectionalfunctions}
\vspace{-0.5cm}
\begin{center}
\begin{tabular}{|p{5cm}|p{11cm}|}
 \hline
{\bf Bidirectional Message Exchanges} & {\bf Function of \emph{Downlink} Communication}  \\\hline
(Data, \emph{ACK/Data}) & To acknowledge confirmed uplink packets and/or send downlink data\\\hline
(LinkCheckReq, \emph{LinkCheckAns}) & To report uplink quality to end devices \\\hline
(\emph{LinkADRReq}, LinkADRAns) & To adapt communication parameters for uplink transmissions\\\hline
(\emph{DutyCycleReq}, DutyCycleAns) & To set transmission duty cycle for end devices \\\hline
(\emph{RXParamSetupReq}, RXParamSetupAns) & To set the communication settings for RX2 window and offset between uplink and RX1 window\\\hline
(\emph{DevStatusReq}, DevStatusAns)& To request battery status and quality of downlink from end devices \\\hline
(\emph{NewChannelReq}, NewChannelAns) & To create/modify settings of the channels to be used for communication\\\hline
(\emph{RXTimingSetupReq}, RXTimingSetupAns) & To set the time offset between the uplink TX and the RX1 window\\\hline
\end{tabular}
\end{center}
\vspace{-6mm}
\end{table*}

With the realization that most studies evaluate \lorawan under uplink traffic only, LoRaWANSim aims to provide support for bidirectional communication by adding downlink. We consider it fundamental for multiple convincing reasons, as described in the introduction.
Table~\ref{tab:bidirectionalfunctions} shows a non-exhaustive list of bidirectional exchanges/handshakes between end devices and gateways. It clearly highlights the importance as well as variety of the functions for which \lorawan relies on downlink. We now explain the actual mechanisms that govern bidirectional communications in LoRaWANSim.

\subsubsection{Downlink data \& acknowledgment traffic}

\begin{figure}
\centering
\vspace{0.1cm}
\includegraphics[scale=0.55,Trim=0mm 30mm 4mm 4mm, clip=true]{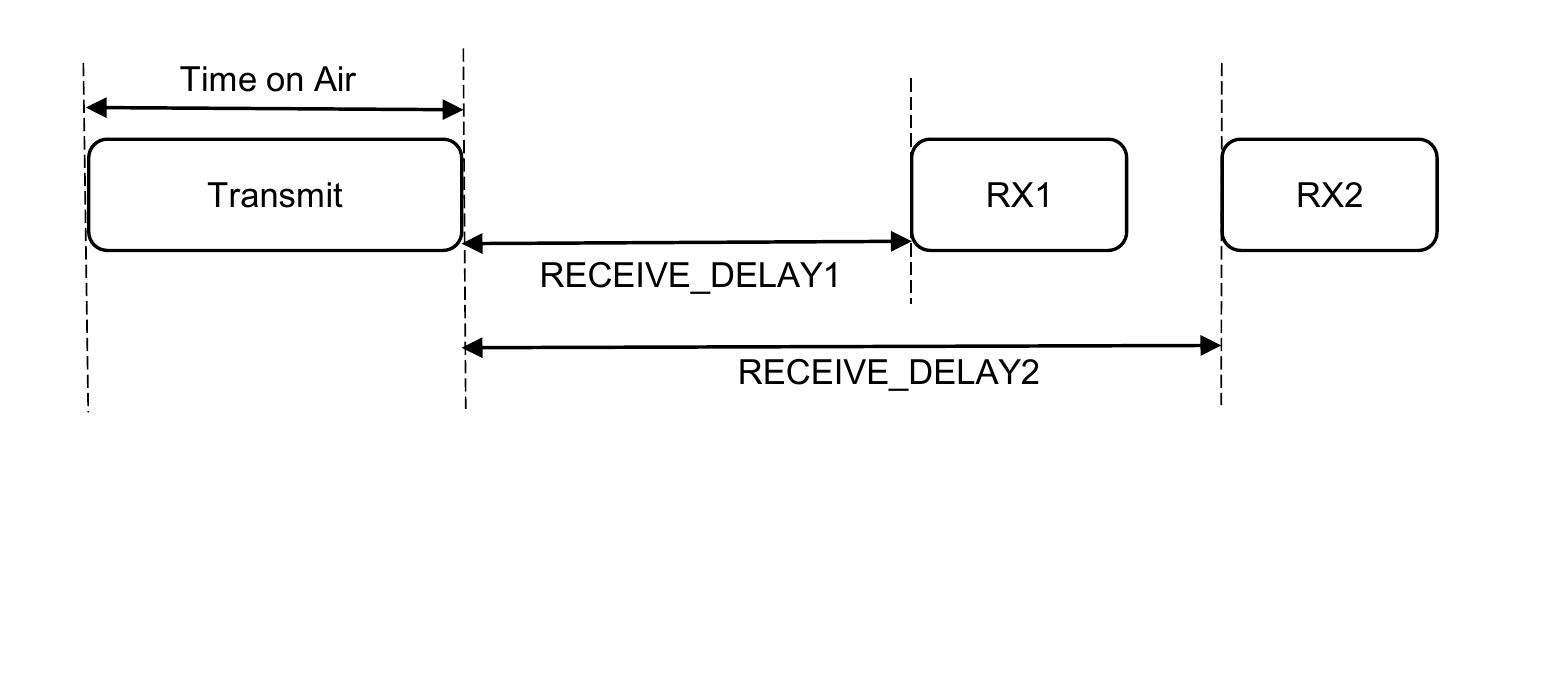}
\caption{Class A end devices open the two windows (RX1 \& RX2) after a transmission to receive an ACK or any other downlink traffic}
\label{fig:classa}
\vspace{-0.3cm}
\end{figure}

We rely on Figure~\ref{fig:classa} to explain the implementation of downlink traffic in LoRaWANSim. The figure shows the timing diagram of a \lorawan Class A end device. The device first transmits its data to a gateway and may request an ACK by setting the acknowledgment bit if desired. It then opens up to two reception windows (RX1 and RX2), giving opportunities to the gateway to send an ACK (if requested) or data (if available) in the downlink. LoRaWANSim implements this and also allows piggybacking ACKs on downlink data traffic. 

By default, the same frequency is used for RX1 as the preceding uplink reception. For RX2, the default frequency, spreading factor and bandwidth are specified to be 869.545 MHz, SF12 and 125 kHz in Europe. Nevertheless, these settings can be programmed individually for each end device. LoRaWANSim allows users to configure the default settings or alternatively use the same data rate as that of RX1. 

\subsubsection{Duty cycle limitation}
To respect the 1\% duty cycle in most European sub-bands, LoRaWANSim ensures that any two consecutive packet transmissions are separated by at least 99x the time-on-air of the first packet. Nothing can be transmitted in this period, including ACKs. This applies to gateways and end devices as per the \lorawan specification.   

\subsubsection{Uplink/downlink collision model}

LoRaWANSim implements a realistic collision model in which downlink and uplink transmissions will not collide even if they happen to overlap in time and frequency and use the same SF. This is because the gateway applies \memo{hardware signal processing techniques (I/Q inversion) when transmitting, which ensure that only end devices} can hear the gateway and vice versa. \memo{This prevents} the problem of the transmission of a node interfering with that of the gateway. \memo{An end device will only receive the packet sent by the gateway, even though it will experience higher RSSI coming from neighboring nodes if it is part of a cluster of nodes far away from a gateway}.

\subsubsection{Retransmission strategy}

\memo{
Retransmissions may follow if a packet is not acknowledged. An acknowledgment may not be received by an end device due to multiple reasons: \emph{ (a)} The confirmed packet requesting the ACK is lost. This may happen due to collisions with other transmissions.\emph{ (b)} The duty cycling regulations prevent the gateway from transmitting. ISM band regulations severely reduce the number of nodes that a gateway can send to. This is by far the most common cause of lost downlink packets as highlighted in Section~\ref{sec:eval}.

When an ACK is lost due to \memo{any reason}, the end device may retransmit the packet multiple times. The \lorawan specification recommends to transmit up to 8 times. If eight consecutive attempts fail, the application should be notified.  

\subsubsection{Data rate adaptation under packet loss}

The \lorawan specification recommends that end devices reduce the data rate every two unsuccessful transmission attempts to achieve more robust uplink connectivity. 
LoRaWANSim supports two options: to either stick to the original data rate for successive attempts or reduce it as per the recommendations. Each new packet is sent at the original data rate in this study.}



\memo{

\section{LoRaWAN in Action: Performance under Bidirectional Traffic}
\label{sec:eval}
}
It is of great interest to assess how a network using LoRaWAN would perform in the presence of downlink traffic in addition to uplink traffic. As mentioned before, existing literature in this area only provides insight for the uplink only scenario. In this paper, LoRaWANSim is used to study system performance under different scenarios with downlink data and ACKs in addition to the uplink traffic. We use the same set of scenarios employed in~\cite{lancasterlorascale}, with the only difference being the \memo{addition} of ACKs/downlink traffic depending on the scenario under study.

Some of the key questions we are attempting to answer by investigating performance in \memo{several} scenarios are:
\begin{itemize}
    \item What is the reduction in network capacity resulting from the introduction of downlink traffic (data and/or ACKs)?
    \item Whilst the LoRaWAN specification recommends 8 transmission attempts, is it really necessary to have such a high number of retransmission attempts?
    \item Each retransmission has an energy cost associated with it. Therefore, how many retransmission attempts might be essential on average to achieve a desired percentage of packet delivery \memo{given the network size and volume of traffic}? Knowledge of such trade-offs can assist \memo{application designers in making appropriate choices} to fulfill the application requirements.
    \item What is the impact on network performance and energy consumption if the percentage of \memo{uplink traffic requiring} ACKs increases?
\end{itemize}

\memo{Several different configurations were considered}:

\begin{itemize}
    \item In one scenario, downlink traffic amounting to a certain percentage of the uplink traffic was \memo{sent}. No acknowledgments were employed in this scenario. As a result, there are no retransmissions. If a packet doesn't get through in the first attempt, the device moves on to the next.
    \item In another scenario, a certain percentage of the uplink traffic requests an ACK from the gateway. There is no downlink data (other than the ACKs) sent by the gateway to the end points. As recommended in the LoRaWAN specification, each node was configured to transmit maximum of 8 times if an ACK is not received.
    \item Finally, a scenario combining \memo{the two is also of interest, i.e., the gateway sends a certain percentage of the uplink traffic as data in the downlink and a certain percentage of end nodes request an ACK from the gateway.}
\end{itemize}

\subsection{Simulation Setup and Performance Metrics}

The topology used in the simulation study comprised of a gateway with nodes randomly distributed around it such that the gateway is reachable. A number of setups with a different number of nodes (from 100 to 5000) in each setup were considered in the simulations. The distance between the gateway and each node is different resulting in each node being capable of supporting different data rates. For example, nodes closer to the gateway could use higher data rates whilst nodes farther from the gateway could only resort to using lower data rates. To simulate a realistic LoRaWAN deployment, each end device was set to use the highest data rate that still allows it to communicate with the gateway. In a real network an ADR mechanism would ensure that such a situation is reached. This minimizes time on air, reducing energy consumption and the chance of collision. The nodes are set to one of the 3 frequencies that the specification requires devices to support in Europe: 868.1 MHz, 868.3 MHz, and 868.5 MHz. These belong to the g1 sub-band, with a duty cycle limit of 1\%. The bandwidth is set to 125 KHz, which is the only setting that allows all spreading factors to be used. Each simulation was run for an equivalent of 57 days and repeated 15 times. The results shown are an average over these, while standard deviations are omitted from the plots as these are too small to be noticeable. 
We use goodput as a network performance metric, which we define as the number of successfully received packets that are not retransmissions divided by the total number of sent packets including retransmissions.

\begin{figure}[!t]
\vspace{2mm}
\centering
\begin{minipage}{3.7cm}
\includegraphics[scale=0.15,Trim=20mm 0mm 0mm 22mm, clip=true]{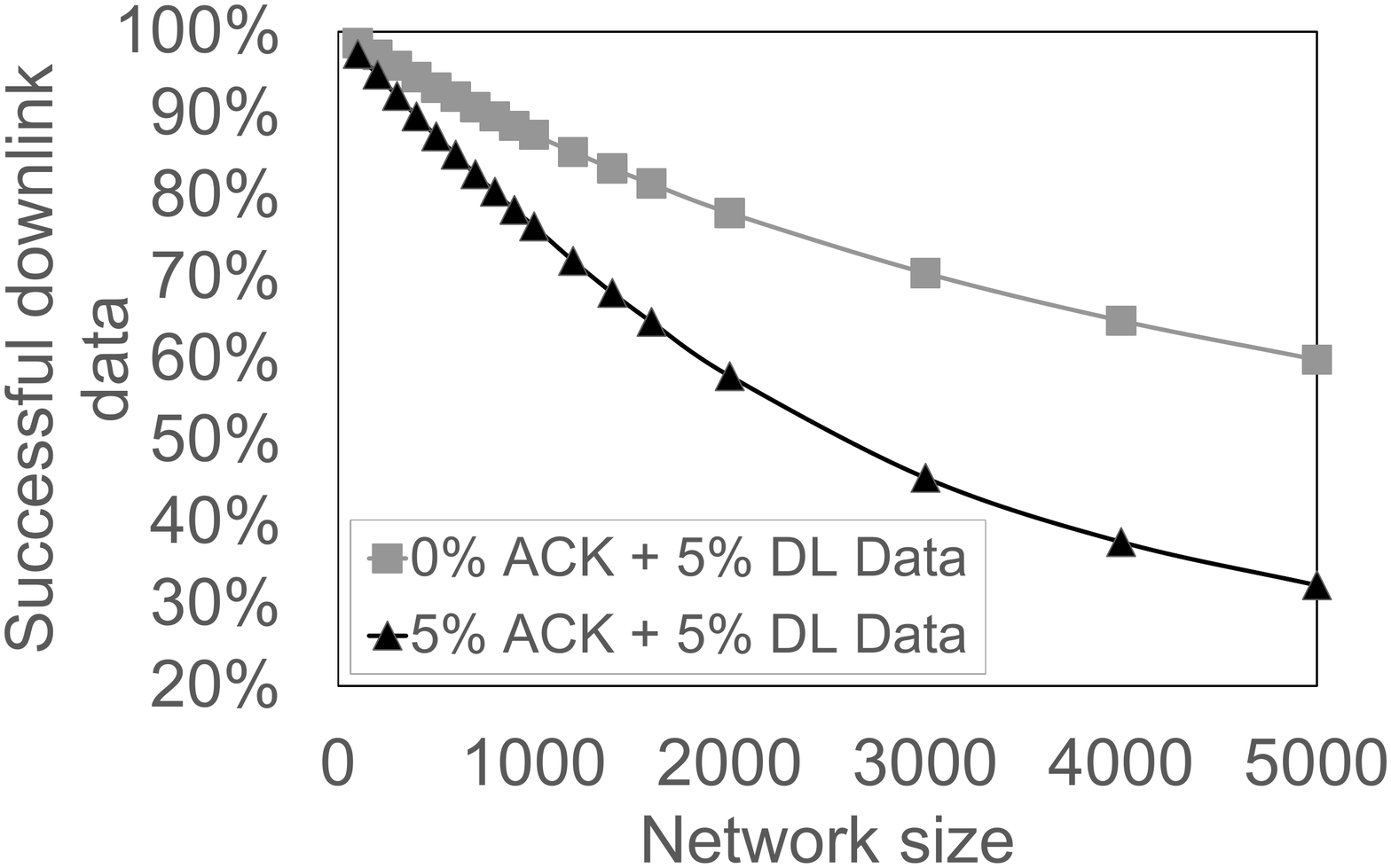}
\vspace{-10mm}
\caption{\camera{Impact of network size on downlink reliability.}}
\label{fig:downlinkreliabilitydc}
\end{minipage}
\hspace{5mm}
\begin{minipage}{3.7cm}
\hspace{-5mm}
\includegraphics[scale=0.15,Trim=20mm 0mm 0mm 22mm, clip=true]{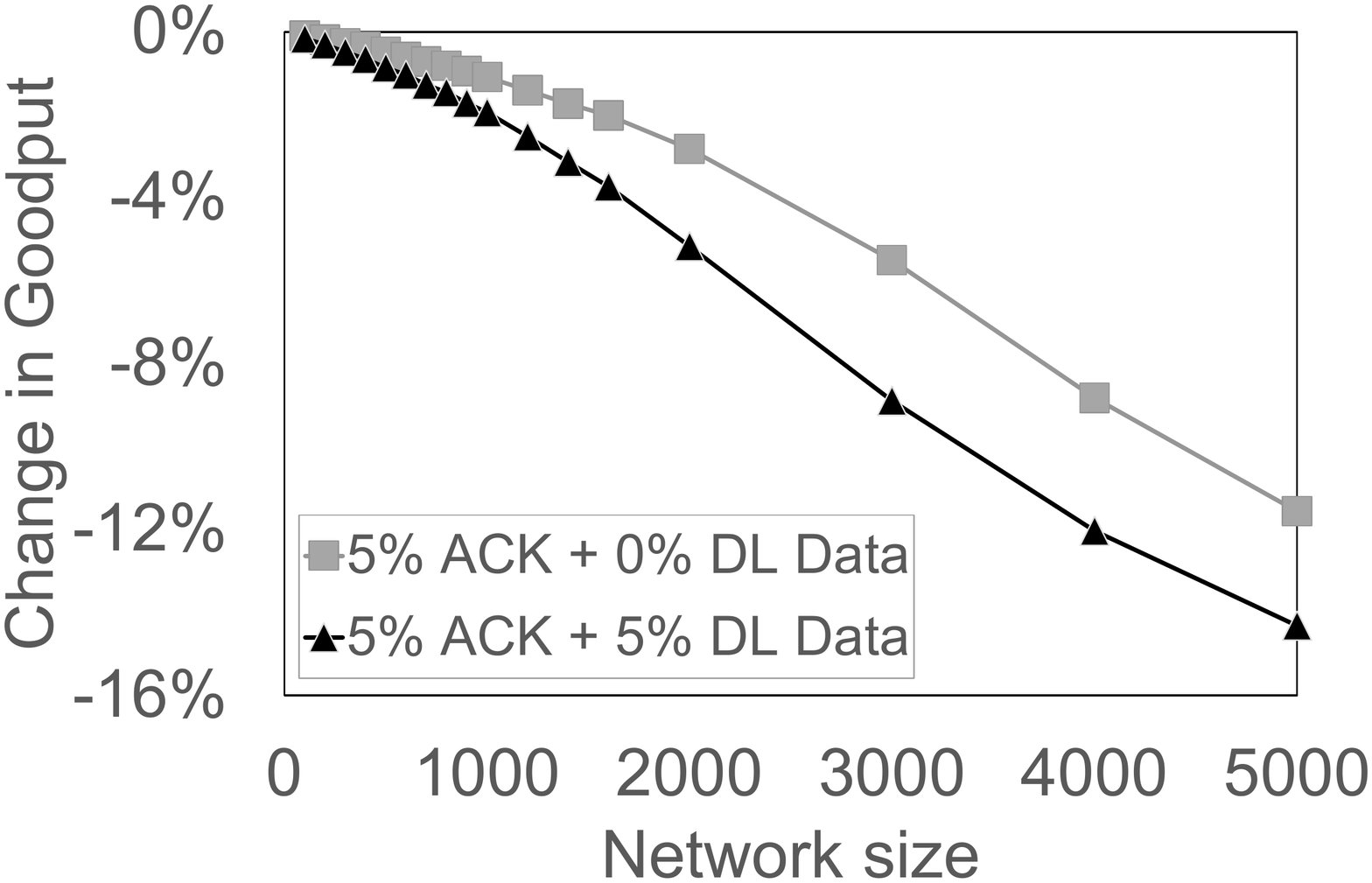}
\vspace{-10mm}
\caption{\camera{Impact of bidirectional traffic on network goodput.}}
\label{fig:capacity5percentdownlink}
\end{minipage}
\vspace{-4mm}
\end{figure}


\subsection{Findings from this study}

\finding{1}{Gateway duty cycle limitations and collisions result in an unreliable downlink in \lorawan}

When communicating to Class A devices, the gateway transmits any downlink traffic and ACKs in the two receive windows (RX1 and RX2) of the end devices as explained in Section \ref{sec:lorawansim}. As the network scales, the gateway is required to transmit more downlink traffic, exhausting its regulatory time on air allocation more often. Each time this happens, the gateway cannot transmit for some time. Figure \ref{fig:downlinkreliabilitydc} shows the percentage of successful downlink transmissions, with and without ACKs. This shows that, especially for large networks, the gateway is often unable to transmit downlink frames. This needs to be taken into account when assessing the suitability of \lorawan for specific applications \memo{or implementing higher layer protocols that require bidirectional communication such as the IPv6 stack}.

\finding{2}{Retransmissions cause a large drop in network goodput} 

Oblivious to the gateway's incapacity to transmit because of its duty cycle, end devices continue their transmissions and expect ACKs against their confirmed messages in one of the two receive windows. Whenever end devices do not hear back from the gateway, they assume loss of their confirmed messages sent over the uplink. This triggers the retransmission of packets. Figure \ref{fig:capacity5percentdownlink} shows the percentage reduction in uplink network goodput when the network is made to acknowledge only 5\% of the total uplink traffic. A reduction in overall goodput of the network is observed for all network sizes. However, the reduction is pronounced for larger network sizes, as evident from Figure \ref{fig:capacity5percentdownlink}. Furthermore, the introduction of downlink data in addition to the ACKs also has a slight impact on the uplink goodput. Whilst one would naturally expect this to happen, the introduction of data messages in the downlink leads to competition with the ACK messages given the duty cycle limitation, resulting in loss of some ACKs thereby triggering retransmissions. These extra retransmissions end up reducing the uplink goodput even further.

\memo{
\finding{3}{Lack of an ACK does not usually mean poor link quality}

We foresee a challenge for \lorawan here. The current \lorawan specification \emph{strongly} recommends that if an end device fails to receive an ACK against its confirmed message after multiple consecutive transmission attempts, it should switch to a lower data rate (higher SF) to achieve better sensitivity. We argue that such strategy may only work if losses at the higher data rate are related to poor link quality and are not due to exhaustion of the duty cycle limit at the gateway. In the latter case, reduction of data rate will require the gateway to send ACKs at the lower data rate, increasing its time on air, further exacerbating the problem.
}


\finding{4}{One size does not fit all: The recommended number of retransmission attempts in the LoRaWAN specification may not be appropriate in all scenarios}
\begin{figure}
\centering
\vspace{2mm}
\includegraphics[scale=0.28,Trim=20mm 20mm 20mm 20mm, clip=true]{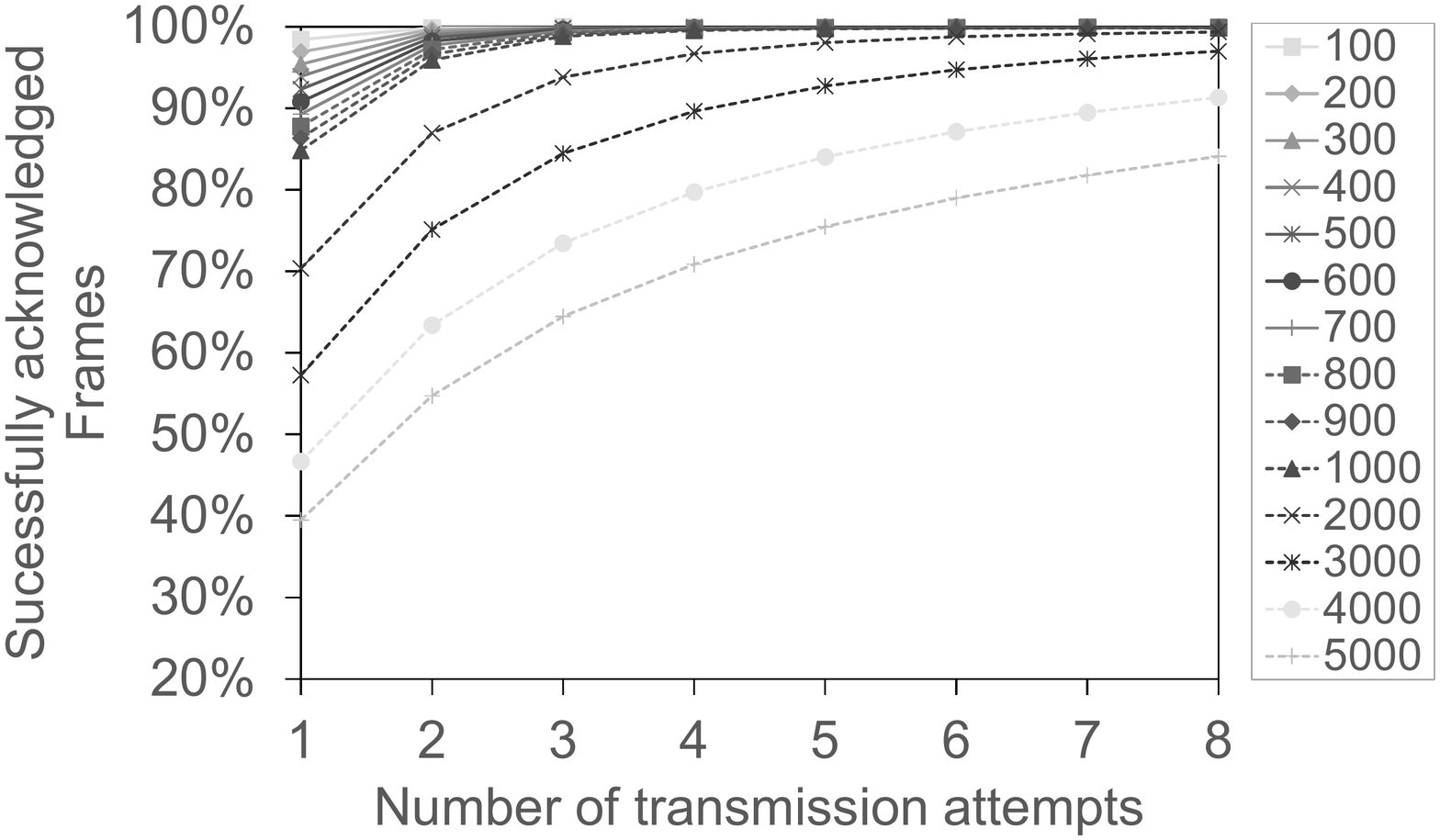}
\vspace{-8mm}
\caption{\camera{Impact of number of retransmissions on network reliability.}}
\label{fig:transmissionretry}
\vspace{-4mm}
\end{figure}

Figure \ref{fig:transmissionretry} shows the percentage of successfully acknowledged messages for a different number of retransmissions for a wide variety of network sizes varying from 100 to 5000. The value of 1 on the X-axis depicts the first transmission attempt with all subsequent values being successive retransmission attempts.
\camera{When the network size is small (less than 600 nodes), 90\% of the messages are acknowledged in the first attempt, with over 95\% of packets being acknowledged for 2 retransmissions for networks of up to 1000 nodes. In contrast, if we look at the 5000-node network, 3 or more transmissions result in a significant improvement in the percentage of acknowledged packets, with about 85\% of acknowledgments being successful after 8 attempts. For a 2000-node network, whether one uses 3 retransmissions or 8 there is at best a marginal improvement in the percentage of packets acknowledged. Therefore, in such cases, the best strategy may be not to rely on retransmissions.}


\begin{figure}
\vspace{-2mm}
\centering
  \subfloat[\camera{Energy Consumption}]{\label{fig:energyretxa} \includegraphics[scale=0.15, Trim=22mm 22mm 22mm 22mm, clip=true]{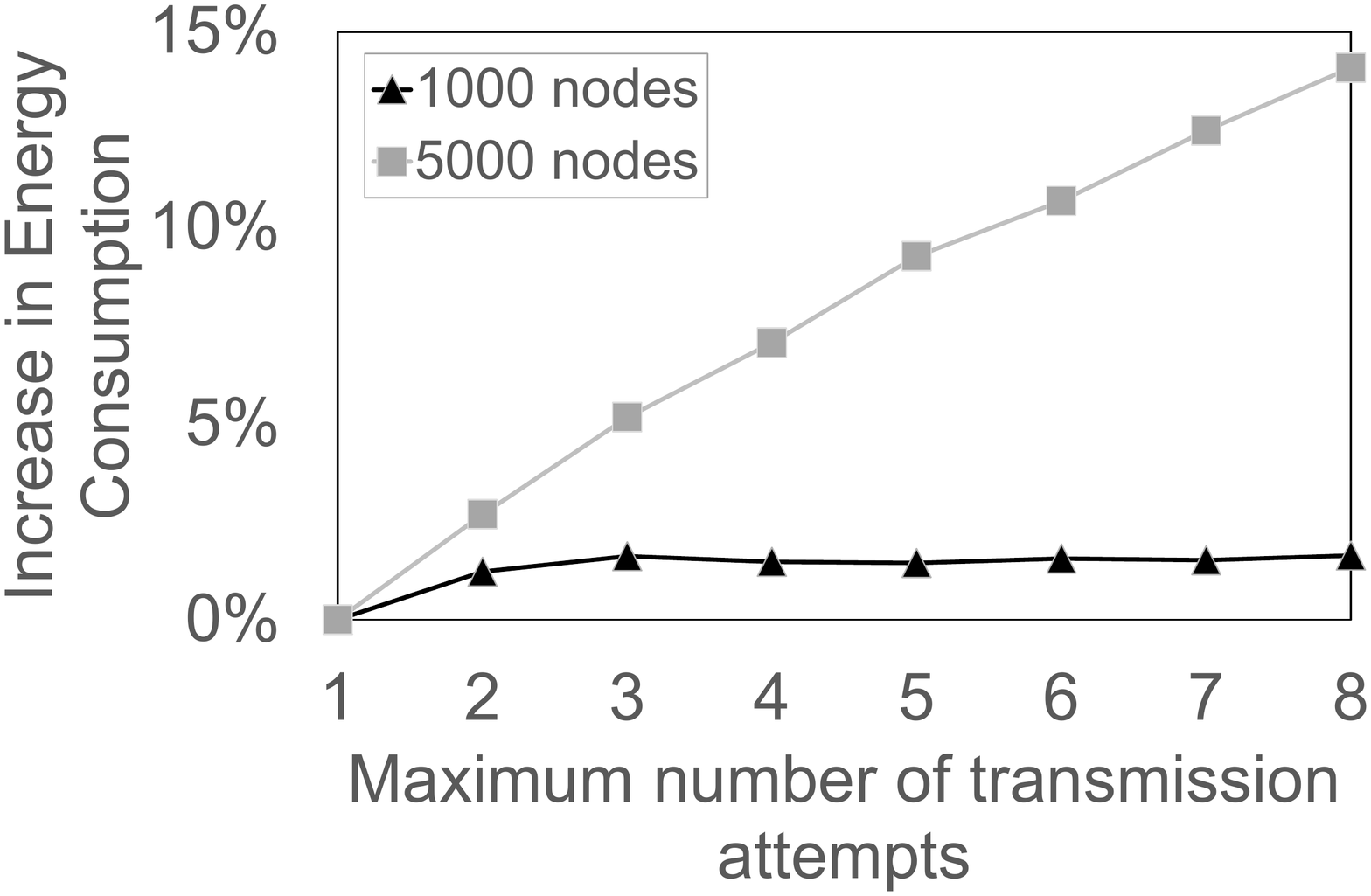}}
  \hspace{2mm}
   \subfloat[\camera{Network Goodput}]{\label{fig:energyretxb}  \includegraphics[scale=0.15, Trim=22mm 22mm 22mm 22mm, clip=true]{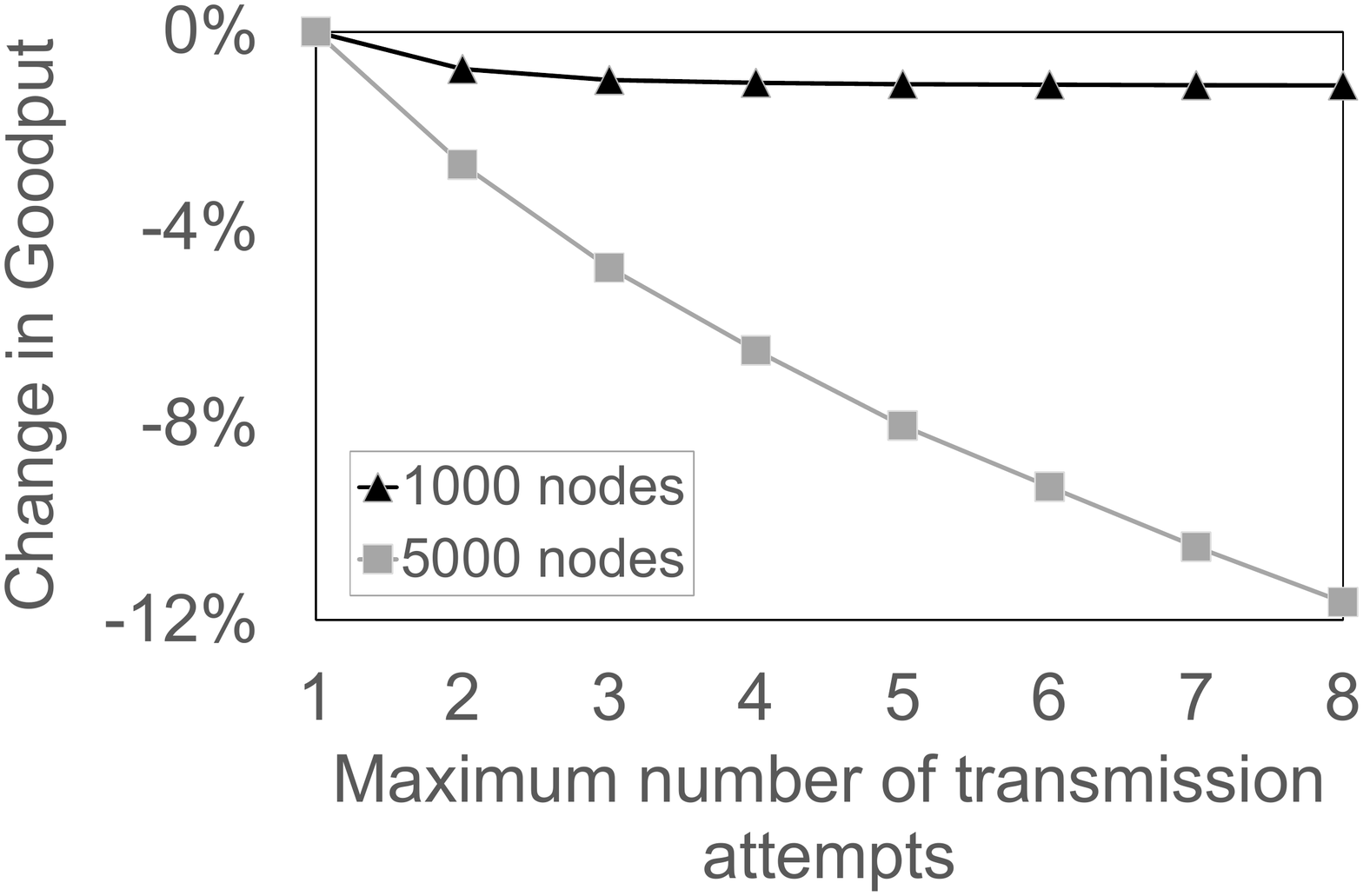}}
  \caption{\camera{Impact of maximum number of retransmissions on \lorawan.}}
\label{fig:energyretx}
\end{figure}

Another interesting issue from an application's perspective is the energy cost of retransmissions and its corresponding impact on network performance. Figure \ref{fig:energyretxa} shows the extra energy consumed (in comparison to the case without retransmissions) for configuring a different number of retransmissions for a 1000 node and 5000 node network respectively. 
Figure \ref{fig:energyretxb} shows the resulting reduction in the performance. As evident from these figures, increasing the number of retransmit attempts \camera{in larger (e.g., 5000 node)} networks not only leads to an increase in battery drain but also reduces network goodput. This is attributed to a combination of effects described earlier (duty cycle limitations at the gateway and the extra contention that drives up collisions).


\finding{5}{LoRaWAN based networks do not scale if a large number of nodes request ACKs}

\begin{figure}
\centering

\vspace{-4mm}
    \subfloat[\camera{Energy Consumption}]{\label{fig:energyackb}  \includegraphics[scale=0.15, Trim=20mm 22mm 22mm 22mm, clip=true]{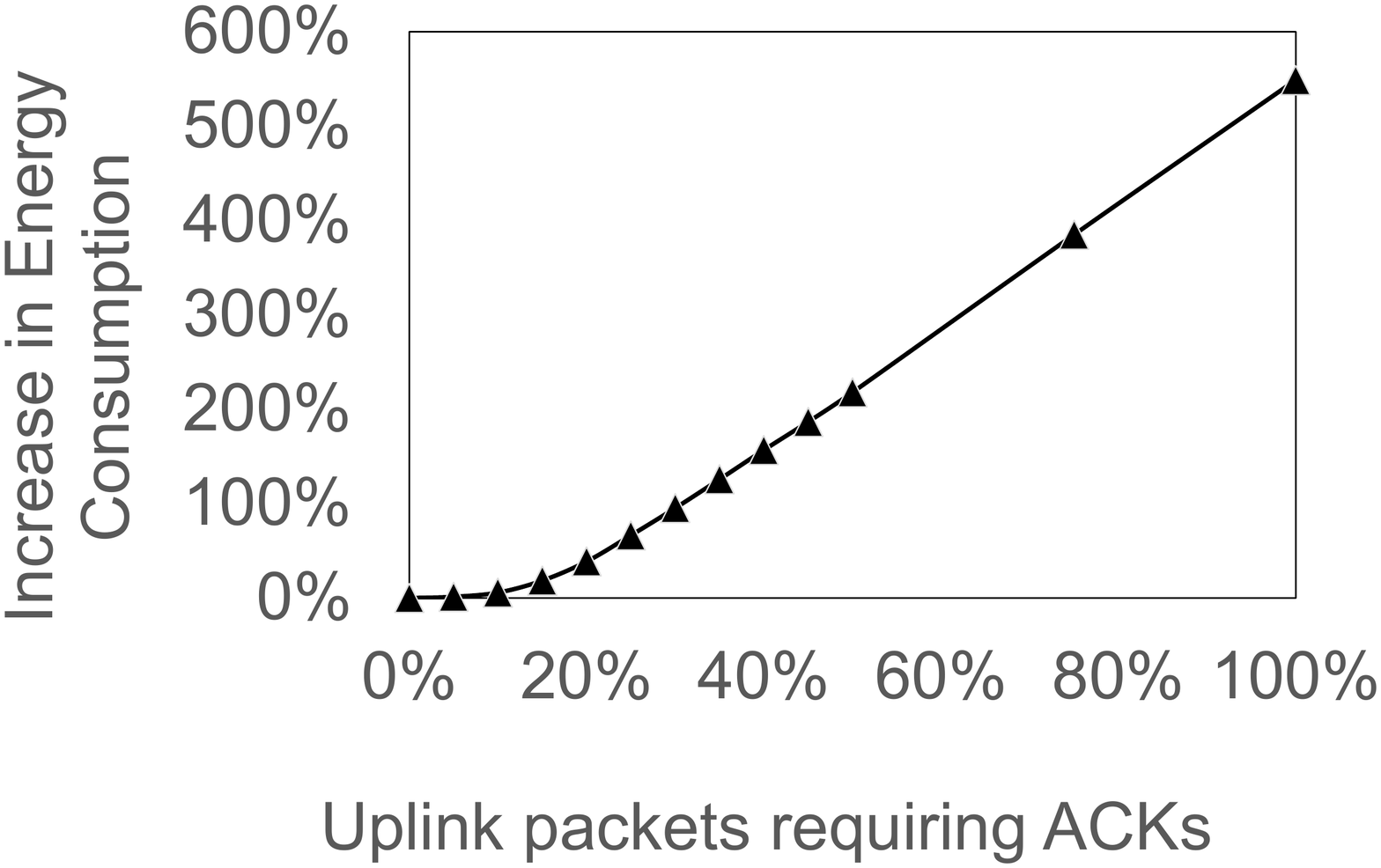}}
    \hspace{2mm}
    \subfloat[\camera{Network Goodput}]{\label{fig:energyacka} \includegraphics[scale=0.15, Trim=29mm 22mm 22mm 22mm, clip=true]{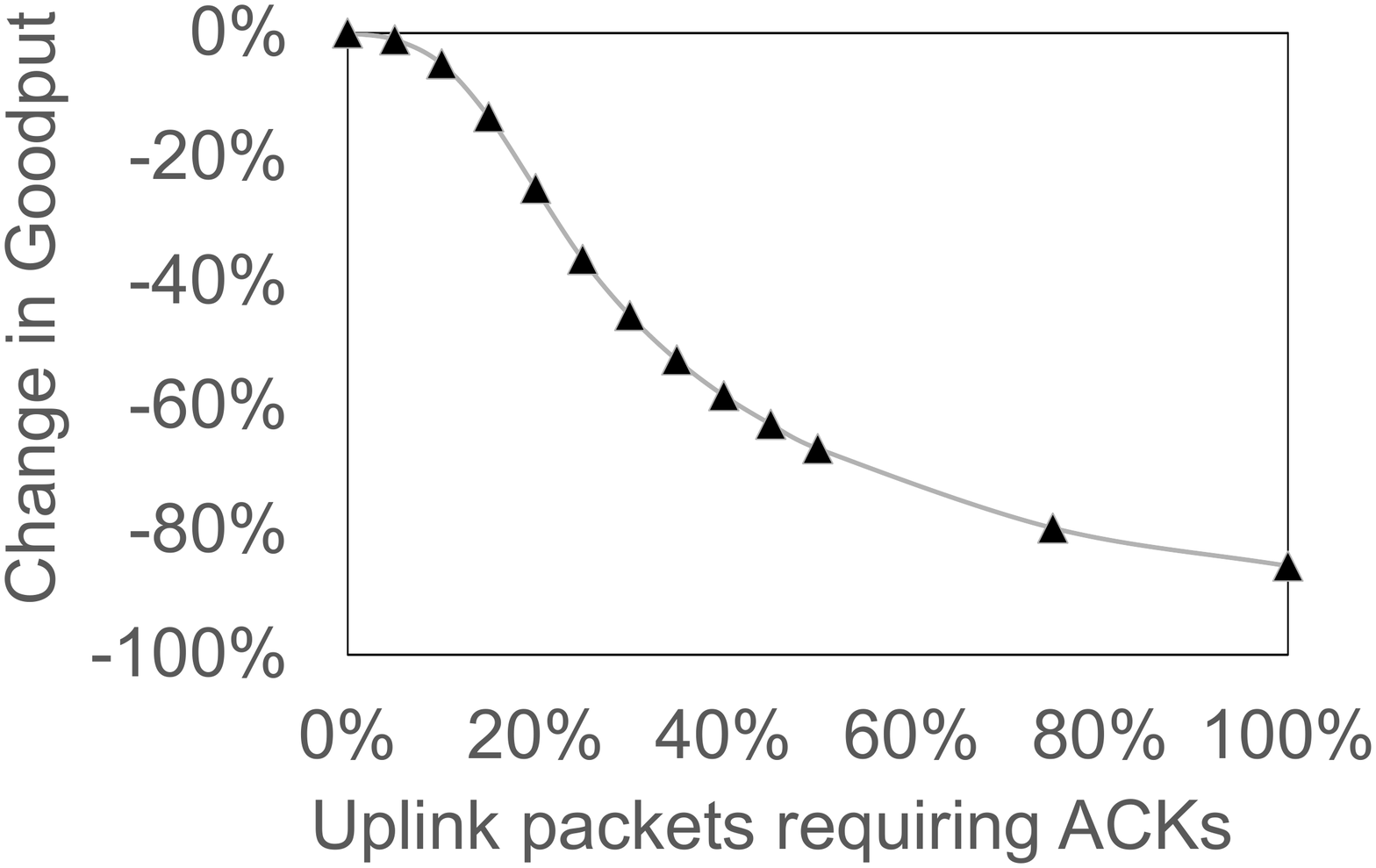}}
  \caption{Impact of percentage of \memo{packets} requiring an ACK on \lorawan. }
\label{fig:energyack}
\vspace{-4mm}
\end{figure}

Figure \ref{fig:energyack} show the effect of different percentage of \memo{packets in the uplink} requiring an ACK. It is evident that as the percentage of \memo{uplink messages} requiring an ACK increase, the network performance severely degrades. With 100\% of \memo{packets} requesting an ACK, the network can barely operate at 15\% of its capacity in comparison to the scenario where no \memo{packets} request an ACK. Clearly, an increase in the percentage of ACKs for larger network sizes will stress test the gateway, which will run out of transmit opportunities often due to duty cycle limitations thereby failing to return a significant number of ACKs. This in turn increases the number of retransmissions, leading to a \camera{considerable} increase in the energy consumption as evident from Figure \ref{fig:energyackb}.

\section{Summary \& Discussion}
\label{sec:summary}

In summary, the key takeaways from this study are:
\begin{itemize}
    \item Whilst \memo{a good first step is made in~\cite{lancasterlorascale}}, the achievable network capacity reported in their work is optimistic. Our work shows that when downlink traffic is introduced in addition to the uplink, goodput drops significantly. This has implications not only on the number of devices that can be supported in a network but also the type of applications that \memo{\lorawan would be suitable for}. Even with a small proportion of traffic requesting ACKs, the achievable goodput in the uplink degrades significantly.
    \item Gateways can become bottlenecks due to the scale of traffic and duty cycle limitations and, therefore, the number of nodes that a gateway \memo{will} serve should be carefully planned before deployment.
    \item The number of transmission attempts recommended in the \lorawan specification (i.e., 8) may not be suitable in all scenarios. As shown by our study, there is a need to consider the scale of the network and energy versus packet delivery trade-offs to ascertain an appropriate \memo{value. This is a decision} the designer is best placed to take depending on the \memo{application requirements}.
    \item LoRaWANSim is a versatile tool which can be invaluable to conduct `what-if analyses' that can aid in pre-deployment analysis. 
\end{itemize}

\section{Conclusion}
\label{sec:conclusion}

In this paper, we presented LoRaWANSim, a discrete event simulator to study a complete network stack comprised of the \lora physical layer and the \lorawan MAC layer. By leveraging the capabilities of LoRaWANSim, we explored the interplay between \lorawan features such as duty cycle limitation and bidirectional communication and provided new insights including trade-offs associated with network scalability, reliability and energy consumption. We revealed that duty cycle limited \lorawan gateways are easily overloaded by downlink traffic. We also showed that these networks do not scale well if many end devices request acknowledgments. 


\balance
\bibliographystyle{abbrv}
\bibliography{bib} 
\end{document}